\documentclass[aip, apl, amsmath, amssymb, reprint]{revtex4-2}

\usepackage{graphicx}
\usepackage{dcolumn}
\usepackage{bm}
\usepackage[T1]{fontenc}
\usepackage{mathptmx}
\usepackage{xcolor}
\usepackage[english]{babel}
\usepackage{upgreek}
\usepackage[hidelinks]{hyperref}

\makeatletter
\def\@email#1#2{%
 \endgroup
 \patchcmd{\titleblock@produce}
  {\frontmatter@RRAPformat}
  {\frontmatter@RRAPformat{\produce@RRAP{*#1\href{mailto:#2}{#2}}}\frontmatter@RRAPformat}
  {}{}
}
\makeatother

\begin{document}

\title{Mutual Inductance Sensing SQUID: Cryogenic microcalorimeter based on mutual inductance readout of superconducting temperature sensors}

\author{J. Zeuner}%
\affiliation{Institute of Micro- and Nanoelectronic Systems (IMS), Karlsruhe Institute of Technology (KIT), Hertzstrasse 16, Building 06.41, D-76187 Karlsruhe, Germany}

\author{C. Schuster}%
\altaffiliation{Present address: Physikalisch-Technische Bundesanstalt (PTB) Braunschweig, Bundesallee 100, 38116 Braunschweig}
\affiliation{Institute of Micro- and Nanoelectronic Systems (IMS), Karlsruhe Institute of Technology (KIT), Hertzstrasse 16, Building 06.41, D-76187 Karlsruhe, Germany}

\author{S. Kempf}%
\email{sebastian.kempf@kit.edu}
\affiliation{Institute of Micro- and Nanoelectronic Systems (IMS), Karlsruhe Institute of Technology (KIT), Hertzstrasse 16, Building 06.41, D-76187 Karlsruhe, Germany}
\affiliation{Institute for Data Processing and Electronics (IPE), Karlsruhe Institute of Technology (KIT), Hermann-von-Helmholtz-Platz 1, Building 242, D-76344 Eggenstein-Leopoldshafen, Germany}

\date{\today}


\begin{abstract}
Superconducting microcalorimeters, such as superconducting transition-edge sensors and magnetic microcalorimeters, have emerged as state-of-the-art detectors for X-ray emission spectroscopy by combining near-unity quantum efficiency with excellent energy resolution. Despite these achievements, their resolving power has not yet reached the level required to rival modern wavelength-dispersive grating or crystal spectrometers. Here, we introduce a next-generation SQUID-based microcalorimeter concept that exploits the strong temperature dependence of the magnetic penetration depth of a superconductor operated close to its critical temperature. The resulting mutual-inductance-based readout enables {\it in situ} tunable signal amplification, while inherently avoiding hysteretic effects that commonly limit superconducting sensors. Experiments with prototype devices demonstrate robust and reproducible operation over a wide temperature range. Based on our measurements and modeling, we project that, using an optimized absorber–sensor combination, an energy resolution below $100\,\mathrm{meV}$ (FWHM) should be achievable for soft X-ray photons with energies below $800\,\mathrm{eV}$. This approach therefore represents a promising pathway towards next-generation cryogenic detectors for high-precision X-ray emission spectroscopy.
\end{abstract}

\maketitle


Cryogenic microcalorimeters have become indispensable tools for various applications, including X-ray emission spectroscopy at synchrotron light sources, ion-beam storage facilities or in laboratory environment~\cite{Friedrich2006, Uhlig2015, Doriese2016, KraftBermuth_2018}. Established microcalorimeter concepts, such as superconducting transition-edge sensors~(TESs)~\cite{Irwin2005,Ullom2015} and magnetic microcalorimeters~(MMCs)~\cite{Fleischmann2005, Bandler2012, Kempf2018}, have demonstrated exceptional performance and combine excellent energy resolution with near-unity quantum efficiency. They operate by detecting the minute temperature rise in an absorber following an X-ray absorption, which is converted into a measurable electrical signal by an ultrasensitive thermometer.

State-of-the-art TESs and MMCs achieve an energy resolution $\Delta E_\mathrm{FWHM}$ of $0.72\,\mathrm{eV}$ for $1.5\,\mathrm{keV}$ photons~\cite{Lee2015} and $1.25\,\mathrm{eV}$ for $5.9\,\mathrm{keV}$ photons~\cite{Krantz2024, Toschi2024}. In terms of resolving power, they thus approach the performance of wavelength-dispersive X-ray spectrometers, yet offer several orders of magnitude higher efficiency. However, despite these remarkable achievements, both TESs and MMCs face inherent challenges, including limited dynamic range, device fabrication complexity, stringent readout noise requirements, or sensor calibration constraints, which currently hinder their routine operation with sub-$1\,\mathrm{eV}$ energy resolution. Among other reasons, these limitations motivate the continued exploration of magnetic penetration depth thermometers (MPTs), exploiting the strong temperature dependence of the magnetic penetration depth of a superconducting thin film near its critical temperature for temperature sensing. MPTs has been already investigated by multiple groups~\cite{Stevenson2013, Dipirro1996, Shirron1993, McDonald1987, Schuster2023}. Building on this work, we propose a new detector concept, the Mutual Inductance Sensing SQUID (MISS). It differs from previous approaches by probing the London penetration depth via a mutual inductance measurement rather than a self‑inductance measurement, and by integrating the readout SQUID directly into the calorimeter.

\begin{figure}
  \includegraphics[width=1.0\columnwidth]{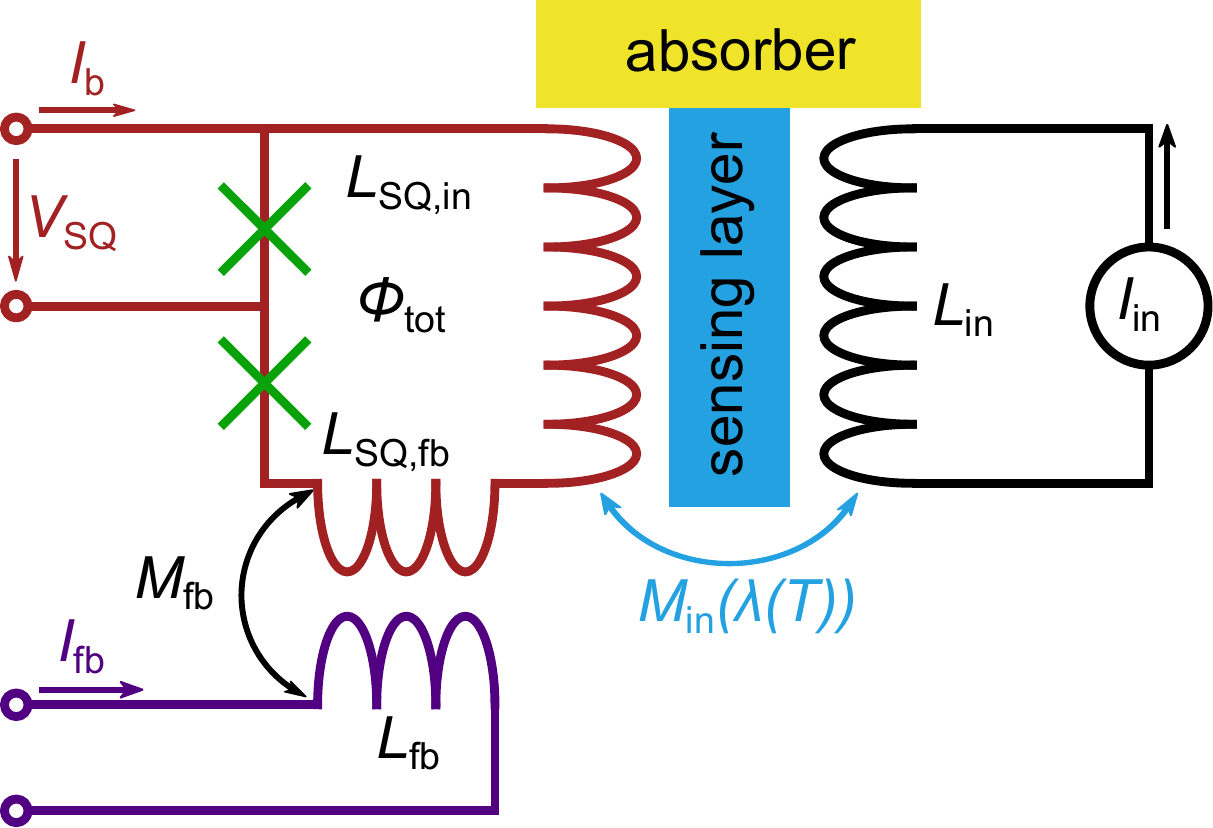}
  \caption{Schematic circuit diagram of a Mutual Inductance Sensing SQUID (MISS). The red-, black-, and purple-colored inductors are made from a superconducting material with critical temperature $T_\mathrm{c}$. The sensing layer (blue) is made from a superconducting material with significantly lower critical temperature, i.e., $T_\mathrm{c,sens} \ll T_\mathrm{c}$, and is in good thermal contact with a suitable X-ray absorber. The device is operated at temperature $T_0 \lesssim T_\mathrm{c,sens}$. A current source injects a constant current $I_\mathrm{in}$ into an input coil with inductance $L_\mathrm{in}$. Owing to the temperature-dependent diamagnetic response of the sensing layer, the mutual inductance $M_\mathrm{in}(\lambda(T))$ between the input coil and the inductively coupled SQUID loop segment with inductance $L_\mathrm{SQ,in}$ becomes temperature dependent. The green crosses indicate resistively shunted Josephson tunnel junctions.}
  \label{fig:LSQ_shematic}
\end{figure}

Figure~\ref{fig:LSQ_shematic} shows a simplified schematic circuit diagram of a Mutual Inductance Sensing SQUID (MISS), highlighting its key components. The central element is an ultrasensitive thermometer that exploits the strong temperature dependence of the magnetic penetration depth of a superconducting thin film operated slightly below its transition temperature $T_\mathrm{c, sens}$. The thermometer is sandwiched between a conventional direct-current superconducting quantum interference device (dc-SQUID) and a superconducting input coil, both fabricated from a superconducting material with critical temperature $T_\mathrm{c} \gg T_\mathrm{c, sens}$. In addition, it is thermally coupled to a suitable particle absorber. The sensing layer is electrically floating, i.e., it has no galvanic contact with either the dc-SQUID or the input coil. Owing to its diamagnetic response, the superconducting sensing layer partially screens the SQUID loop from the magnetic flux generated by the input coil. As a consequence, the mutual inductance between the the input coil and the SQUID loop depends on the screening properties of the sensing layer. Since this screening is governed by the magnetic penetration depth $\lambda(T)$ of the sensing film, the mutual inductance becomes temperature dependent, i.e., $M_\mathrm{in} = M_\mathrm{in}\left( \lambda \left( T \right) \right)$. By appropriate design of the sensing layer geometry and the choice of material, a very large temperature responsivity can be achieved, enabling the realization of highly sensitive microcalorimeters.

The dc-SQUID used for mutual inductance sensing consists of a superconducting loop with total inductance $L_\mathrm{SQ} = L_\mathrm{SQ,in} + L_\mathrm{SQ,fb}$, which is interrupted by two Josephson tunnel junctions, each shunted by a resistor $R$. Magnetic flux is coupled into the SQUID loop either via the input coil with inductance $L_\mathrm{in}$ or via a feedback coil with inductance $L_\mathrm{fb}$. The feedback coil enables flux-locked loop (FLL) SQUID operation as well as the application of a static flux bias for device characterization. The SQUID may be operated under either voltage or current bias. Using a constant bias current $I_\mathrm{b}$, the SQUID voltage $V_\mathrm{SQ}$ exhibits a periodic dependence on the magnetic flux threading the SQUID loop.

The mutual couplings between SQUID loop to both the feedback coil and the input coil are quantified by the mutual inductances $M_\mathrm{in}$ and $M_\mathrm{fb}$, respectively. For the feedback coil, the mutual inductance $M_\mathrm{fb} = k_\mathrm{fb} \sqrt{L_\mathrm{fb} L_\mathrm{SQ,fb}}$, with $k_\mathrm{fb}$ denoting the geometric coupling factor, is temperature independent. In contrast, the mutual inductance $M_\mathrm{in}(\lambda(T)) = k_\mathrm{in}(\lambda(T)) \sqrt{L_\mathrm{in} L_\mathrm{SQ,in}}$ is temperature dependent due to diamagnetic response of the superconducting sensing layer. Here, $k_\mathrm{in}$ denotes the geometric coupling factor that depends on the magnetic penetration depth $\lambda(T)$ of the superconducting screening layer separating the SQUID loop and the input coil. Close to the transition temperature $T_\mathrm{c,sens}$, the magnetic penetration depth $\lambda(T)$ exhibits a strong temperature dependence. Consequently, a temperature change $\Delta T$ induces a change in SQUID output voltage $\Delta V_\mathrm{SQ}$ according to
\begin{equation}
    \Delta V_\mathrm{SQ} = \frac{\partial V_\mathrm{SQ}}{\partial T} \Delta T
        = \frac{\partial V_\mathrm{SQ}}{\partial \Phi} \frac{\partial M_\mathrm{in}}{\partial T}I_\mathrm{in}\Delta T
    \label{eq:SQUID_Voltage}
\end{equation} 
with the gain coefficient $\partial V_\mathrm{SQ}/\partial T$ and $I_\mathrm{in}$ denoting the constant current running through the input coil.

\begin{figure*}[t]
  \includegraphics[width=\textwidth]{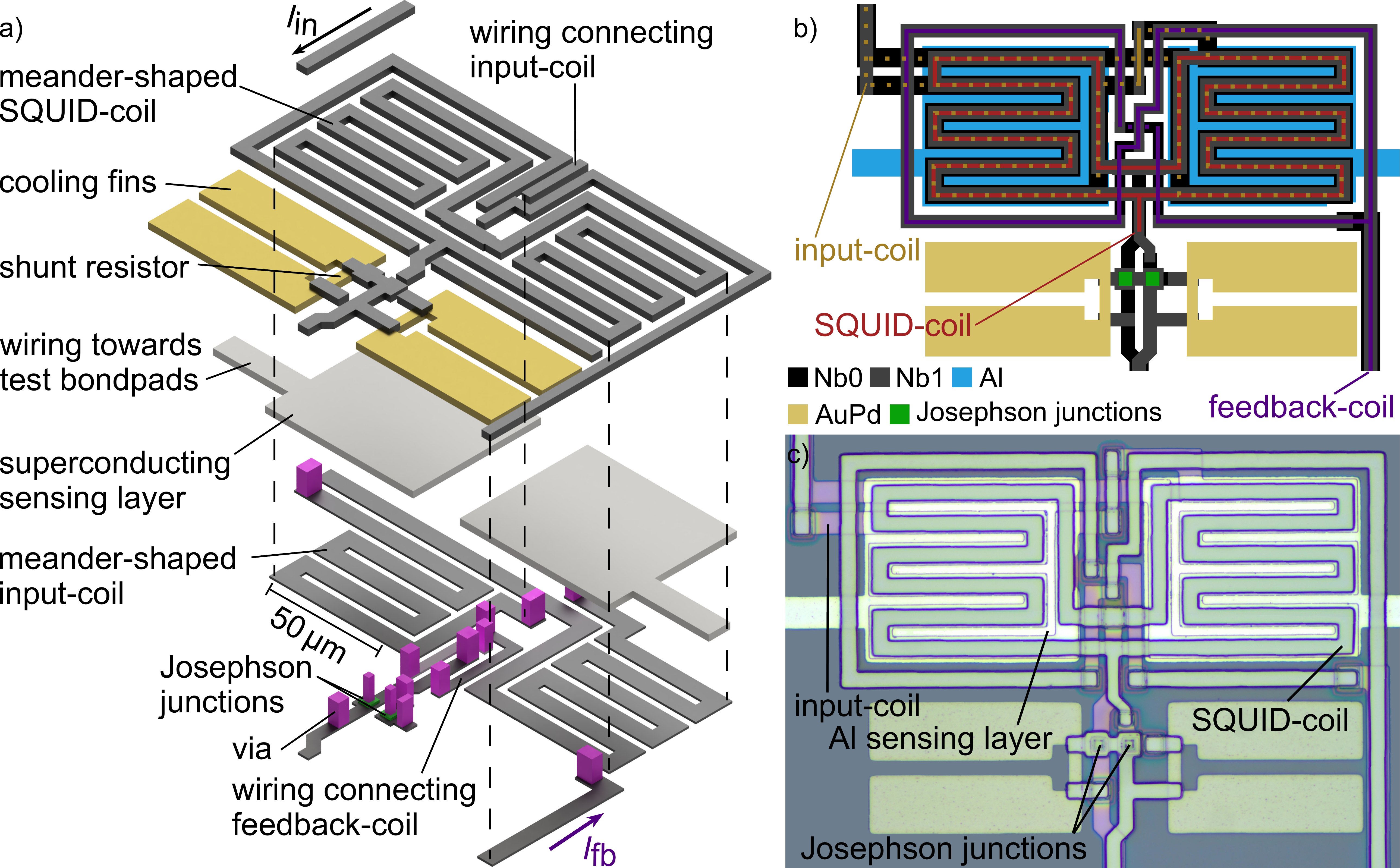}
  \caption{(a) Exploded view drawing, (b) simplified layout drawing, and (c) micrograph of the prototype Mutual Inductance Sensing SQUID (MISS). The sensing layer (silver in (a) and (c), blue in (b)), is made from Al ($T_\mathrm{c,Al} \simeq 1.2\,\mathrm{K}$), while the other superconducting device components are made from Nb ($T_\mathrm{c} \simeq 9.0\,\mathrm{K}$). In (b), the input and flux biasing coil as well as the SQUID loop are traced in yellow, purple and red, respectively.}
  \label{fig:design}
\end{figure*}

To demonstrate the feasibility of the proposed microcalorimeter concept and to experimentally assess its achievable performance, we designed, fabricated, and comprehensively characterized a simple prototype MISS (see figure~\ref{fig:design}). The SQUID inductance is formed by two meander-shaped coils arranged in a first-order parallel gradiometer configuration. The superconducting sensing layer, fabricated using standard process parameters for the aluminum layer of our Josephson junctions~\cite{Adam2024}, consists of a patterned $\simeq 125\,\mathrm{nm}$ thick aluminum film sandwiched between the input coil and the SQUID loop. It is electrically floating and has no electrical contact with any other device component. Although aluminum is not an optimal choice for a practical detector with sub-eV energy resolution, due to its high transition temperature, we selected this material in order to defer the implementation of a dedicated low-$T_\mathrm{c,sens}$ sensor material deposition process. For the same reasons, we omitted a particle absorber. Similar to the device reported by Krantz~{\it et al.}~\cite{Krantz2024}, the feedback coil surrounds the SQUID loop. This configuration, however, proved to be suboptimal, as the feedback coil is also partially screened by the superconducting sensing layer, rendering the corresponding mutual inductance temperature dependent. A full functional microcalorimeter incorporating a sensing layer with $50\text{-}100\,\mathrm{mK}$ transition temperature, a suitable X-ray absorber, and a feedback coil geometry unaffected by the screening layer will therefore be the subject of future work. As the prototype MISS was not optimized for performance, we fabricated a second prototype device (MISSv2), in which we increased the area of the sensing layer by $33\,\%$. As a result, the sensing layer extends beyond the SQUID loop.

\begin{figure*}[t]
    \centering
    \includegraphics[width=\textwidth]{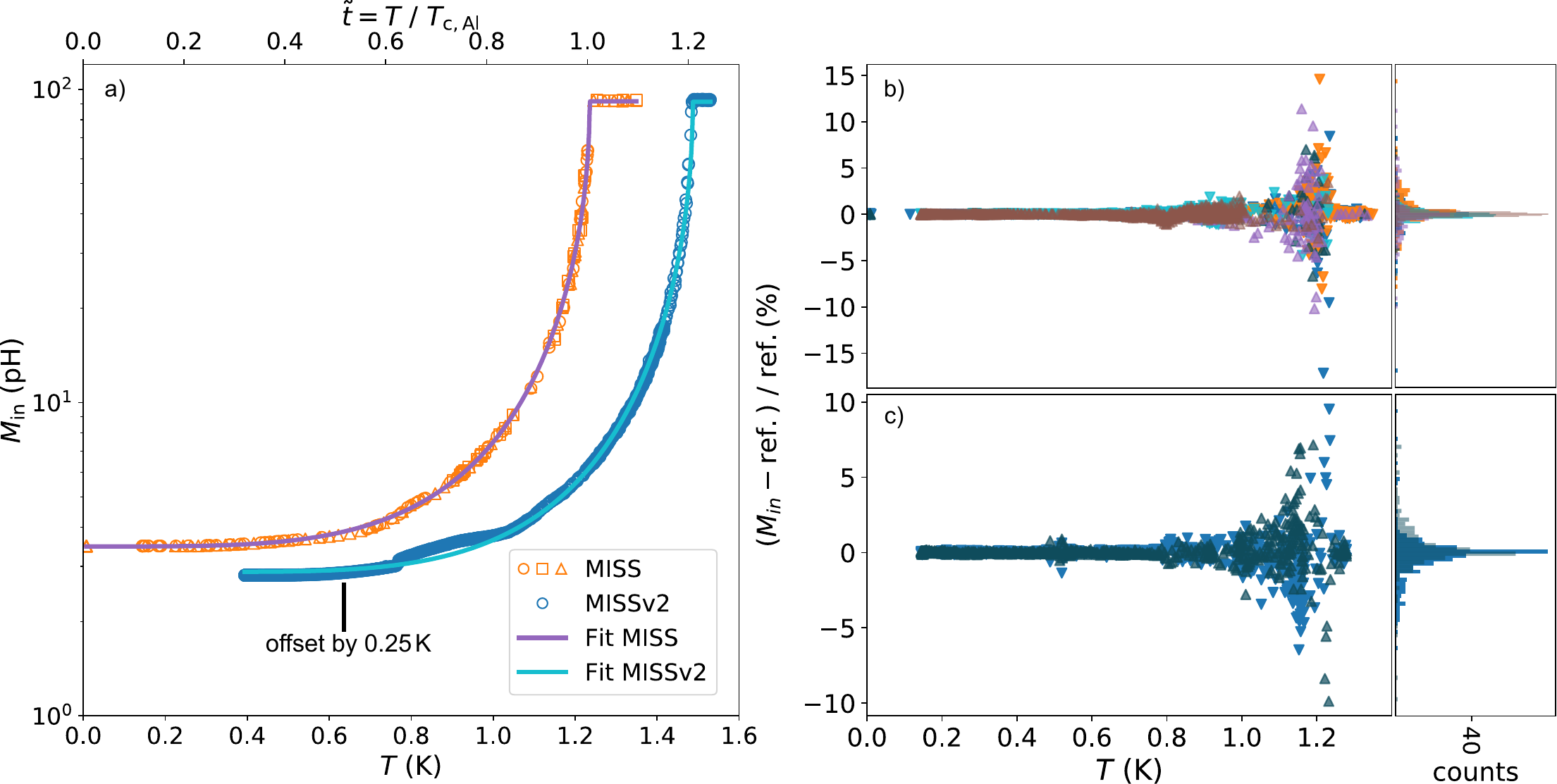}
    \caption{(a) Measured mutual inductance $M_\mathrm{in}$ as a function of temperature $T$ (bottom axis) and reduced temperature $\tilde{t} = T/T_{\mathrm{c,Al}}$ for both prototype MISS devices. For clarity, the MISSv2 data and corresponding fit are horizontally shifted by $0.25\,\mathrm{K}$, and only a random subset of data points is shown for the original MISS. For the original variant, three consecutive temperature cycles were recorded, indicated by different symbols. Their perfect overlap demonstrates the absence of hysteresis. Solid lines represent fits to the measured data.
    (b) Deviation of individual measurements from the averaged response function for the original MISS, separated into increasing temperature ($\blacktriangledown$, various colors) and decreasing temperature ($\blacktriangle$, various colors). A histogram of the deviation data above $700\,\mathrm{mK}$ is also shown.
    (c) Same analysis as in (b), but for MISSv2. Here, the averaged function used for comparison with the decreasing temperature data ($\blacktriangle$, dark blue) is derived from the increasing temperature data ($\blacktriangledown$, light blue), and vice versa.}
    \label{fig:Mutualinductance}
\end{figure*}

We comprehensively characterized both prototype MISSs in a $^{3}$He/$^{4}$He dilution refrigerator. By adjusting the cryostat operating conditions, the mixing chamber temperature was swept between $7\,\mathrm{mK}$ and $1.3\,\mathrm{K}$. The MISS was operated in current-bias mode using a direct-coupled high-speed dc SQUID electronics~\cite{Drung2006}. The same electronics were used both to operate the SQUID and to inject the drive current into the input coil. The drive current had a triangular waveform with a repetition rate of $20\,\mathrm{Hz}$ and a peak-to-peak amplitude of $1\,\mathrm{mA}$. Prior to digitization, the SQUID output voltage was amplified by the dc SQUID electronics with a gain of 2000. The resulting SQUID output voltage $V_\mathrm{SQ}$ as well as the drive current were sampled for $320\,\mathrm{ms}$ using an digital oscilloscope. This measurement cycle was repeated at one-minute intervals while the temperature was slowly varied. The mutual inductance $M_\mathrm{in}$ was determined offline from the periodicity of the acquired SQUID flux-voltage response: For each rising and falling edge of the drive current triangular signal, the periodicity of $V_\mathrm{SQ}$ was determined by the position of first peak of the autocorrelation function. Using the corresponding current change $\Delta I_\mathrm{in}$, the mutual inductance was calculated through $\frac{1}{M_\mathrm{in}} = \Delta I_\mathrm{in} / \Phi_0$. The data points shown in figure~\ref{fig:Mutualinductance} represent the mutual inductance obtained from averaging the different mutual inductance values determined within a single measurement. To estimate the uncertainty in $\Delta I_\mathrm{in}$, both the full width at half maximum (FWHM) of the first peak in the autocorrelation function and the sample standard deviation over all current edges were evaluated. Each point exhibit a combined uncertainty that is negligible compared to the symbol size in figure~\ref{fig:Mutualinductance}a). Special attention was given to identify hysteretic behavior. To this end, three independent measurement cycles were performed while minimizing thermal gradients between the MISS and the mixing chamber plate. Future work will involve higher drive current amplitudes and repetition rates, which are not supported by the present setup.

Figure~\ref{fig:Mutualinductance}a) shows the measured temperature dependence $M_\mathrm{in}(T)$ of the mutual inductance for both prototype devices. For the original MISS, three consecutive temperature cycles were performed; however, for improved visibility, only a representative selection of the data points from all cycles is shown in figure~\ref{fig:Mutualinductance}a). For MISSv2, only a single cycle was recorded due to limited cryostat access. Moreover, the respective data are shown with a horizontal offset of $0.25\,\mathrm{K}$ for clarity. For the original MISS, all cycles overlap, indicating no hysteresis. This was confirmed by constructing a reference curve from the averaged interpolations of the three measurement cycles that was compared to data separated into rising and falling temperature branches. Figure~\ref{fig:Mutualinductance}b) shows the resulting deviation of the individual curves from the average for $T > 700,\mathrm{mK}$, giving not hint for a systematic deviation and hence hysteresis.
For MISSv2, sufficient data density enables interpolation of the separated branches; the reference was therefore constructed using the interpolation of the opposite branch. Figure~\ref{fig:Mutualinductance}c) shows the corresponding deviation analysis.

At temperatures $T>1.24\,\mathrm{K}$, the sensing layer is in the normal-conducting state. Consequently, it has a negligible influence on the magnetic coupling between the SQUID loop and the input coil, and the mutual inductance approaches a value of $M_\mathrm{in} = 93\,\mathrm{pH}$, which corresponds to that of an otherwise identical device without a sensing layer but with the same interlayer spacing. Below its transition temperature, the sensing layer enters the Meissner state, and strongly expels external magnetic fields. This magnetic screening leads to a pronounced reduction of the mutual inductance $M_\mathrm{in}$, reaching near-maximum screening at temperatures below approximately $500\,\mathrm{mK}$. We observe a residual coupling of $M_\mathrm{in} < 4\,\mathrm{pH}$ even at the base temperature of the cryostat. This is expected as our screening layer is not infinite and has only a thickness of few magnetic penetration depths. Although the enlarged sensing layer of our second prototype device has little effect on the screening properties at $T > 1.1\,\mathrm{K}$, it reduces the residual coupling from $3.2\,\mathrm{pH}$ to $2.6\,\mathrm{pH}$. The step-like features in the temperature range between $0.6\,\mathrm{K}$ and $0.9\,\mathrm{K}$ are not captured by our model, and may be attributed to local phase variations within the sensing layer, arising from inhomogeneities in film properties, temperature distribution, and/or stray magnetic fields.

To quantitatively describe our data, we used the following approach. We considered the decay of the magnetic field inside a superconductor due to the Meissner effect. For an infinitely large superconductor, filling the half space $x > 0$ and subjected to a magnetic field pointing in $z$-direction, the decrease of the $z$-component of the magnetic field in $x$-direction is given by
\begin{equation}
    B_z(x)=B_z(0)\cdot \exp(-x/\lambda(T)),
\end{equation}
where $\lambda$ is the London penetration depth. If we consider a thin superconducting layer, for which the thickness is in the order of the penetration depth, the magnetic field exiting the superconductor is finite. Therefore, the magnetic flux in a coil placed behind the superconducting layer is given by $\Phi(x)=\int B(x)dA$, where $A$ is the area of the coil. Consequently, the mutual inductance between two coils separated by a thin superconducting layer is expected to follow the dependence
\begin{equation}
    M_1(T)=M(0)\cdot \exp(-x/\lambda(T))
\end{equation}
with
\begin{equation}
\lambda(T) =
    \begin{cases}
        \frac{\lambda_0}{\sqrt{1-(\frac{T}{T_\mathrm{c}})^4}} &\text{for } T<T_\mathrm{c} \\
        \infty & \text{otherwise}
    \end{cases}.
\end{equation}
As our prototype device is based on a superconducting screening layer with finite extension, we added a constant term $M_\mathrm{off,1}$ to the fit function $M_\mathrm{fit,1}(T)=M_1(T)+M_\mathrm{off,1}$ to include residual fields. Although the resulting fit, shown in figure~\ref{fig:Mutualinductance}a), describes the measurement data well, some clear deviations are visible. An in-depth investigation of these deviations is planned for the future.
          
\begin{figure*}
    \centering
    \includegraphics[width=1.0\textwidth]{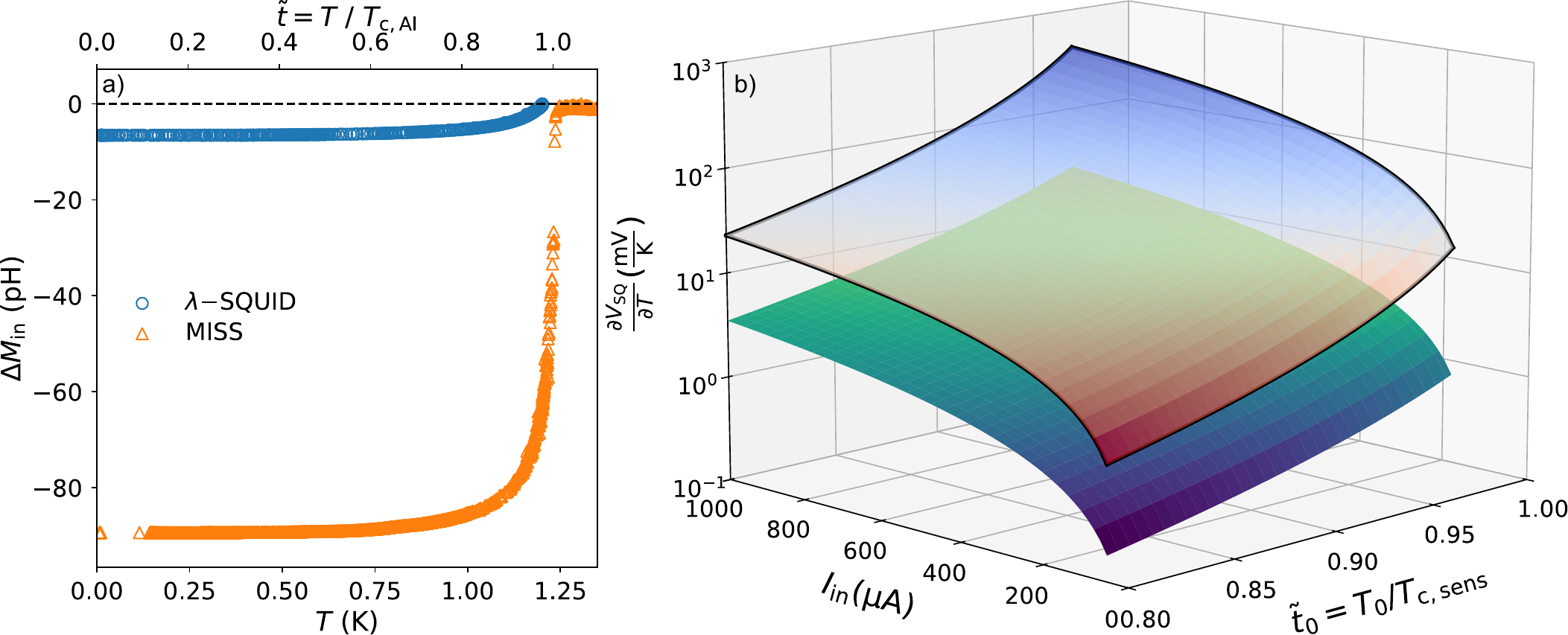}
    \caption{a) Comparison of the change in mutual inductance as function of temperature (bottom axis) and reduced temperature $\tilde{t} = T/T_\mathrm{c,Al}$ for a $\lambda$-SQUID (green) and our MISS (red). The significantly larger change in mutual inductance of the MISS compared to the $\lambda$-SQUID is clearly visible. b) Comparison of the gain factor of the MISS (opaque with contour) and the $\lambda$-SQUID (colored), both calculated from the derivative of the fitted measurement data and assuming a $T_\mathrm{c}$ of $50\,\mathrm{mK}$ of the temperature-sensitive circuit element. The strong dependence on $I_\mathrm{in}$ and the operation temperature $T_0$ is visible as the gain factor changes over two orders of magnitude.}
    \label{fig:gain}
\end{figure*}

At this point we want to compare the sensitivity of our MISS with that of the $\lambda$-SQUID~\cite{Schuster2023}, a superconducting microcalorimeter that likewise exploits the temperature dependence of the magnetic penetration depth of a superconductor operated close to its critical temperature. Similar to the MISS, the $\lambda$-SQUID relies on the temperature-dependent mutual inductance $M_\mathrm{in}$ between the  SQUID loop and a superconducting input coil. In contrast to the MISS, however, the $\lambda$-SQUID senses the change in penetration depth not via a separate, electrically floating sensing layer, but through a change of the effective inductance of a section of the SQUID loop caused by current redistribution. Figure~\ref{fig:gain} illustrates the substantially larger variation of the mutual inductance, $\Delta M_{\mathrm{in}}\simeq 93\,\mathrm{pH}$ for the MISS and $\Delta M_{\mathrm{in}}\simeq 6\,\mathrm{pH}$ for the $\lambda$-SQUID. In addition, the MISS offer a simpler fabrication process, as the sensing layer is electrically isolated from all other circuit elements and hence does not form an integral part of the SQUID loop, avoiding, for instance, vertical interconnect accesses (vias) between different materials in the SQUID loop.

The gain coefficient $\partial V_\mathrm{SQ}/\partial T$ (see equation~\ref{eq:SQUID_Voltage} and figure~\ref{fig:gain}b)) is an important figure of merit for assessing the performance of the MISS as a microcalorimeter. It relates a temperature change $\Delta T$ upon energy deposition to the corresponding change in the SQUID output voltage $\Delta V_\mathrm{SQ}$. From equation~\ref{eq:SQUID_Voltage}, we identify three key parameters that determine the gain coefficient: a high SQUID gain $\partial V_\mathrm{SQ} / \partial\Phi$, a large temperature sensitivity of the mutual inductance ${\partial M_\mathrm{in}}/{\partial T}$ and a large input current $I_\mathrm{in}$. Increasing any of these parameters enhances the gain coefficient, and consequently the sensitivity of the MISS. Moreover, the explicit dependence of the gain factor on the bias current $I_\mathrm{in}$ enables \textit{in situ} tuning of the detector gain, even during event detection. This provides additional operational flexibility, for example, to optimally match the detector output to the dynamic range of digitizers in the readout chain.

\begin{figure}
    \centering
    \includegraphics[width=1.0\columnwidth]{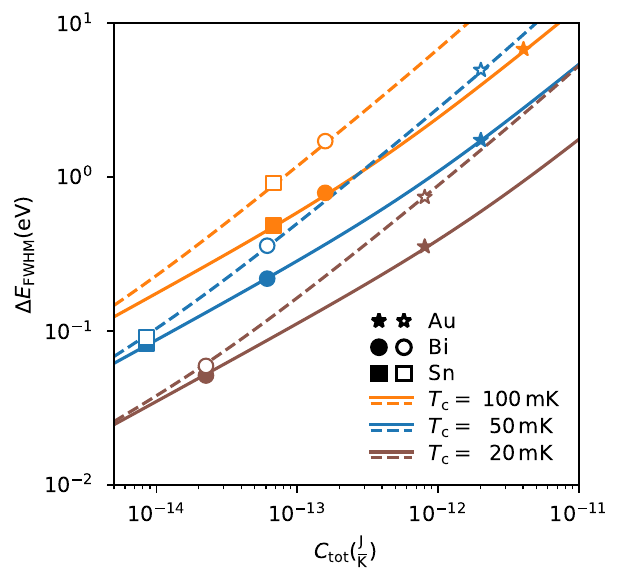}
    \caption{Estimated achievable energy resolution as a function of total heat capacity for a MISS (solid lines) and a $\lambda$-SQUID (dashed lines). Orange, blue and brown represent critical temperatures of $100\,\mathrm{mK}$, $50\,\mathrm{mK}$, and $20\,\mathrm{mK}$ of the temperature-sensitive circuit element. Symbols represent the achievable energy when using a $250\,\upmu\mathrm{m} \times 250\,\upmu\mathrm{m}$ particle absorber made of gold, bismuth, or tin whose thickness was chosen to achieve a stopping power exceeding $99.99\,\%$ for soft X-ray photons photons. filled symbols refer to a MISS, open symbols to a $\lambda$-SQUID.}
    \label{fig:Energy}
\end{figure}

To estimate the achievable energy resolution $\Delta E_\mathrm{FWHM}$ of a MISS employing a sensing layer $T_\mathrm{c,sens} < 100\,\mathrm{mK}$, we follow the approach introduced by Schuster~{\it et. al.}~\cite{Schuster2023} for the $\lambda$-SQUID. We assume that the temperature dependence of the mutual inductance $M_\mathrm{in}(\tilde{t})$ is governed solely by the reduced temperature $\tilde{t} = T/T_\mathrm{c,sens}$. Accordingly,
\begin{equation}
    \frac{\partial M}{\partial T}=\frac{\partial M}{\partial \tilde{t}} \frac{\partial \tilde{t}}{\partial T}.
\end{equation}
This assumption is well justified, as the magnetic penetration depth of superconductors exhibits the same reduced-temperature scaling. To estimate the achievable energy resolution for low-$T_\mathrm{c,sens}$ sensor materials based on our measured data (see figure~\ref{fig:gain}), we analyzed the relevant noise contributions. Assuming the noise contribution of the readout electronics to be negligible by designing a proper dc-SQUID and SQUID-based preamplifier, two dominant noise sources remain: intrinsic SQUID noise $\sqrt{S_\mathrm{E,SQ}}$ and thermodynamic energy fluctuations $\sqrt{S_\mathrm{E, TD}}$ among absorber, sensor, and heat bath~\cite{McCammon2005}. The intrinsic SQUID noise contribution is given by
\begin{equation}
    \sqrt{S_\mathrm{E, SQ}}
        = \sqrt{S_\mathrm{VV}} \left(\frac{\partial V_\mathrm{SQ}}{\partial T}
        \frac{\partial T}{\partial E}\right)^{-1},
    \label{eq:SQUID-noise}
\end{equation}
where $\sqrt{S_\mathrm{VV}}$ denotes the SQUID voltage noise, $\partial V_\mathrm{SQ}/\partial T$ the gain coefficient (see equation~\ref{eq:SQUID_Voltage}), and $\partial T/\partial E = 1/C_\mathrm{det}$ the inverse total heat capacity of the MISS, including the particle absorber. For an optimized SQUID~\cite{Tesche1977}, the SQUID voltage noise is $S_\mathrm{VV}=18 k_\mathrm{B} TR$ and the gain coefficient can be expressed as 
\begin{equation}
    \frac{\partial V_\mathrm{SQ}}{\partial T} =  \frac{\partial V_\mathrm{SQ}}{\partial \Phi} \frac{\partial M_\mathrm{in}}{\partial \tilde{t}} \frac{I_\mathrm{in}}{T_\mathrm{c,sens}}.
\end{equation}
Here, $\partial M_\mathrm{in}/\partial \tilde{t}$ is obtained from the fit of the measurement data and $\partial V_\mathrm{SQ}/\partial \Phi = R/L_\mathrm{SQ}$ with $\beta_L \simeq \beta_c \simeq 1$. The contribution of thermodynamic energy fluctuations $\sqrt{S_\mathrm{E, TD}}$ can be estimated using thermal properties of the detector~\cite{Fleischmann2005}:
\begin{equation}
    S_\mathrm{E,TD} = k_\mathrm{B} C_\mathrm{sens} T^2 \left[ \frac{4  \left(1 - \beta \right) \tau_0}{1 + \left( 2 \pi \tau_0 f \right)^2} + \frac{4 \beta \tau_1}{1 + \left(2 \pi \tau_1 f\right)^2} \right].
    \label{eq:Thermal-noise}
\end{equation}
Here, $C_\mathrm{sens}$ is the heat capacity of the sensor, $\beta = C_\mathrm{sens} / C_\mathrm{det}$ its ratio to the total heat capacity of the MISS, and $\tau_0$ and $\tau_1$ denote the signal rise and decay time. The latter are set by adjusting the thermal conductance between absorber, sensor and heat bath. The achievable energy resolution $\Delta E_\mathrm{FWHM}$ can then readily be estimated by 
\begin{equation}
    \Delta E_\mathrm{FWHM} = 2 \sqrt{2 \ln 2} \left[ \int\limits_0^\infty  \frac{\left| p(f) \right|^2 }{S_\mathrm{E,TD}(f) + S_\mathrm{E,SQ}(f)} \mathrm{d}f \right]^{-1/2}
\label{eq:energyres}
\end{equation}
with the detector responsivity
\begin{equation}
    \left| p(f) \right| = \frac{2 \beta \tau_1}{\sqrt{1 + \left( 2 \pi \tau_0 f \right)^2} \sqrt{1 + \left( 2 \pi \tau_1 f \right)^2}}.
\end{equation}
Figure~\ref{fig:Energy} compares the achievable energy resolution for a MISS and a $\lambda$-SQUID as a function of the total heat capacity $C_\mathrm{det}$ of the detector. For this comparison, we assume $\tau_0=1\,\upmu\mathrm{s}$ ,$\tau_1=1\,\mathrm{ms}$, and $\beta=1/2$ as well as a conservative gain factor set by a small input current of $I_\mathrm{in} = 500\,\upmu\mathrm{A}$ and an operation temperature of $T_0=0.9 T_\mathrm{c,sens}$. In addition, we highlight three representative detectors that employ the common absorber materials gold, bismuth and tin. The absorber size is set to $250\,\upmu\mathrm{m} \times 250\,\upmu\mathrm{m}$. For each material, we choose the absorber thickness to achieve a stopping power exceeding $99.99\,\%$ for soft X-ray photons up to $1\,\mathrm{keV}$. This results in a thickness of $5\,\upmu\mathrm{m}$ for gold, $8.6\,\upmu\mathrm{m}$ for bismuth and $50\,\upmu\mathrm{m}$ for tin. Under otherwise identical operating conditions, the MISS clearly outperforms the $\lambda$-SQUID. 
For a detector employing a bismuth absorber and a sensing-layer critical temperature of  $T_\mathrm{c,sens} = 20\,\mathrm{mK}$, an energy resolution well below $100\,\mathrm{meV}$ appears feasible. However, it should be noted that in  this case the dynamic range is limited to X-ray photons with energies up to $800\,\mathrm{eV}$ to not saturate the detector. For higher photon energies the necessity of decreased operation temperatures could be partially compensated for by an increase of $I_\mathrm{in}$. Alternatively, operation at temperatures close to $100\,\mathrm{mK}$ with an estimated resolution of approximately $500\,\mathrm{meV}$ would already constitute a significant performance benchmark. 

Our model assumes an intrinsic SQUID flux noise for the SQUID at $100\,\mathrm{mK}$ of $0.13\,\mathrm{\mu \Phi_0/\sqrt{Hz}}$. Due to the strong inverse scaling of the SQUID noise contribution with the total detector heat capacity, reducing $C_\mathrm{det}$ drives the MISS performance increasingly close to the thermodynamic noise limit. In addition, the superconducting sensing layer is expected to exhibit a heat capacity comparable to, or lower than, that of state-of-the-art magnetic microcalorimeters~\cite{Krantz2024}. As a result, MISS detectors are particularly well suited for applications requiring exceptional energy resolution at low particle energies. Finally, we note that the MISS generally allows for two distinct modes of operation. In the first mode, the detector is operated at a temperature close to $T_\mathrm{c,sens}$ and with a small input current $I_\mathrm{in}$. In this regime, the signal is dominated by the large temperature-induced change in the mutual inductance, resulting in a high gain coefficient. At the same time, the dynamic range is limited, and the strong temperature dependence of the sensing-layer heat capacity leads to pronounced nonlinearity. Nevertheless, this mode is expected to provide the lowest achievable energy resolution, and no hysteretic behavior was observed in our measurements. The second mode corresponds to operation at temperatures significantly below $T_\mathrm{c,sens}$. In this regime, the reduced temperature sensitivity of the mutual inductance $\partial M_\mathrm{in}/\partial T$ can be compensated by increasing the input current $I_\mathrm{in}$, thereby maintaining a gain coefficient comparable to that of the first mode. The maximum input current is limited by the ampacity of the input coil and the critical magnetic field of the sensing layer. A more detailed investigation of this operating mode with a sensing layer with $T_\mathrm{c} \ll 200\,\mathrm{mK}$ will be the subject of future work. This comes with further challenges, namely in the integration of such sensing layers, including parasitic thermal coupling between the SQUID shunt resistors and the sensor layer, as well as the coupling of a suitable absorber to the sensing layer, whose heat capacity shows a strong temperature dependence in the vicinity of $T_\mathrm{c}$.

In summary, we have presented a novel SQUID-based superconducting microcalorimeter with in situ tunable gain. The design exploits the steep temperature dependence of the magnetic penetration depth near the superconducting critical temperature ($T_\mathrm{c}$), which modulates the mutual inductance between the input coil and the SQUID loop, while allowing real-time adjustment of the input current. We have designed, fabricated, and thoroughly characterized prototype devices using an Al-based sensor layer to map the temperature-dependent variation of the mutual inductance. Importantly, no evidence of hysteresis is observed in the MISS. Together with results from the $\lambda$-SQUID~\cite{Schuster2023}, this indicates that the proposed detector concept does not suffer from hysteresis, a common limitation in superconducting microcalorimeters. Despite remaining challenges—particularly in the thermal management of the detector—the extrapolated energy resolution below $100\,\mathrm{meV}$ for soft X-ray photon energies below $800\,\mathrm{eV}$ for an optimized absorber sensor combination is highly promising.


\begin{acknowledgments}
We would like to thank A.~Stassen for support during device fabrication. We acknowledge the financial support by the KIT Center of Elementary Particle and Astroparticle Physics (KCETA). Furthermore, C.~Schuster acknowledges financial support by the Karlsruhe School of Elementary Particle and Astroparticle Physics: Science and Technology (KSETA). This work was partially funded by the Deutsche Forschungsgemeinschaft (DFG, German Research Foundation) – Projektnummer (project number) 467785074.
\end{acknowledgments}


\section*{Author Declarations}

\subsection*{Conflict of Interest}
The authors have no conflicts to disclose.


\subsection*{Author Contributions}
{\bf Jodok Zeuner:} Conceptualization (equal); Formal Analysis (lead); Investigation (lead); Software (equal), Visualization (lead); Writing – original draft (lead); Writing/Review \& Editing (equal). 
{\bf Constantin Schuster:} Conceptualization (equal); Formal Analysis (supporting); Investigation (supporting); Software (equal); Writing/Review \& Editing (equal). 
{\bf Sebastian Kempf:} Conceptualization (equal); Formal Analysis (supporting); Funding Acquisition (lead); Investigation (supporting); Project Administration (lead); Resources (lead); Supervision (lead); Visualization (supporting); Writing – original draft (supporting); Writing/Review \& Editing (equal).


\section*{Data Availability Statement}
The data that support the findings of this study are available from the corresponding author upon reasonable request.


\bibliography{bibliography}

\end{document}